\documentclass[12pt]{article}
\usepackage[esperanto]{babel}
\usepackage{epsfig}

\newcommand{\bea}{\begin{eqnarray}}
\newcommand{\eea}{\end{eqnarray}}
\newcommand{\dd}{\mathrm{d}}

\begin{document}
\selectlanguage{esperanto}

\title{\bf Relativeca Dopplera efekto ^ce unuforme akcelata movo -- II}  
\author{F.M. Paiva \\ 
{\small Departamento de F\'\i sica, U.E. Humait\'a II, Col\'egio Pedro II} \\
{\small Rua Humait\'a 80, 22261-040  Rio de Janeiro-RJ, Brasil; fmpaiva@cbpf.br} 
\vspace{.7ex} \\
A.F.F. Teixeira \\
{\small Centro Brasileiro de Pesquisas F\'\i sicas} \\
{\small 22290-180 Rio de Janeiro-RJ, Brasil; teixeira@cbpf.br}} 

\maketitle 

\begin{abstract} 
Da^urigante~\cite{PaivaTeixeira2007}, luma fonto de unukolora radiado ^ce rekta movo ^ce konstanta propra akcelo pasas preter restanta observanto. ^Ce la special-relativeco, ni priskribas la observatan Doppleran efikon. Ni anka^u priskribas la interesan nekontinuan efikon se trapaso okazas.   
 \\ - - - - - - - - - - - \\ 
Extending~\cite{PaivaTeixeira2007}, a light source of monochromatic radiation, in rectilinear motion under constant proper acceleration, passes near an observer at rest. In the context of special relativity, we describe the observed Doppler effect. We describe also the interesting discontinuous effect when riding through occurs. An English version of this article is available.   
\end{abstract}

\section{Enkonduko}                         \label{secEnkonduko}
Cita^jo~\cite{PaivaTeixeira2007} studis Doppleran efikon de lumo eligita el restanta fonto, vidata per akcelata observanto kiu pasas preter a^u tra la fonto. ^Ci tie ni inversas tiun sistemon. Nun, fonto de unukolora radiado ^ce rekta movo ^ce konstanta propra akcelo pasas preter a^u tra restanta observanto. La frekvenco $\nu$ de la lumo eligata estas konstanta; tamen, la observata frekvenco (t.e., la koloro) estas $\nu_{obs}$, malsama ol $\nu$. Tio estas nomata Dopplera efiko. La proporcio $D=\nu_{obs}/\nu$, nomata Dopplera faktoro, estas studata ^ci tie. 

\begin{figure}[t]                                              
\hfill
\epsfig{file=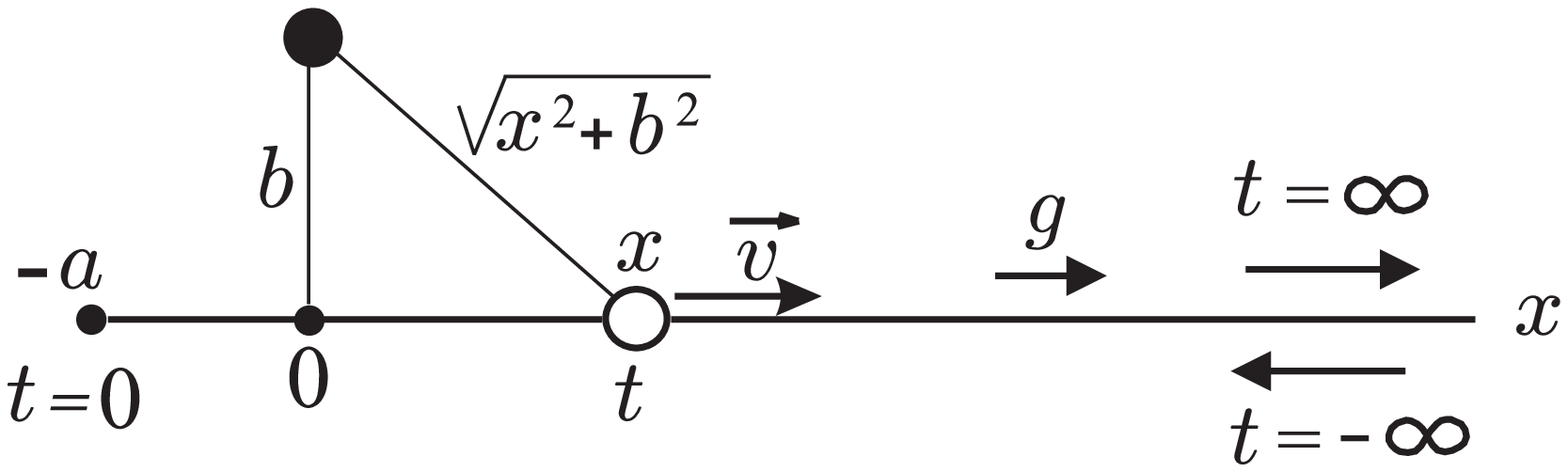,width=8cm}
\hfill
\epsfig{file=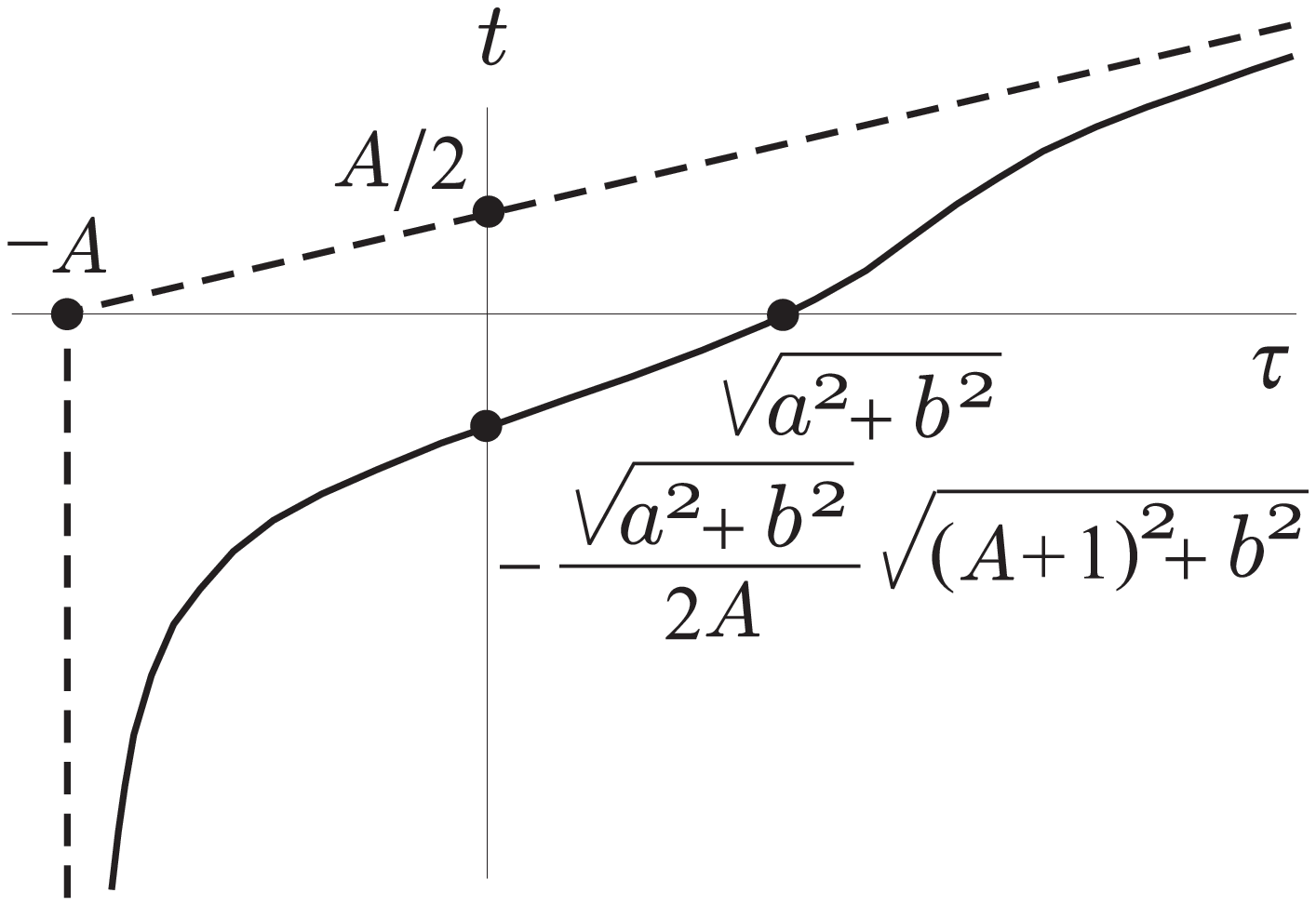,width=6cm}
\hfill \mbox{}
\caption{La sistemo. {\bf \ref{Figura1}.a)} Observanto (nigra sfereto) estas fiksa distance $b$ de akso $x$. Luma fonto (blanka sfereto) estas iranta de $x\!=\!-a$ al $x\!=\!\infty$, kun konstanta propra akcelo $g$. Anta^ue ^gi venis el $x\!=\!\infty$ ($t\!=\!-\infty$), kun konstanta propra malakcelo $g$, ^gis $x\!=\!-a$ ($t\!=\!0$). 
\newline
{\bf \ref{Figura1}.b)} La movi^ganta fonto eligas signalon en la momento  $t$ de la inercia sistemo $S$, kaj la restanta observanto ricevas tiun signalon en la momento $\tau$, ekv.~(\ref{taut}). Vidu la asimptotojn $t\!=\!(\tau+A)/2$ kaj $\tau\!=\!-A$. Vidu anka^u ke ne estas ricevo en tempo anta^u $\tau\!=\!-A$.} 
\label{Figura1} 
\end{figure}

Esti^gu inercia referenco sistemo $S=\{t,\vec{x}\}$. El nefinia loko $x=\infty$, je nefinia estinta momento $t=-\infty$, luma fonto ekvenis al origino $x=0$, per la akso $x$. ^Gia komenca rapido estis $-c$, kaj ^gi estas malakcelata per konstanta propra akcelo $g$; tiel ^gi pasas la originon, poste momente restas ^ce $x=-a$ je $t=0$, kaj tuj reiras al nefinio $x=\infty$ kun sama akcelo. En $x=0$, distance $b$ el akso $x$, estas restanta observanto, kiel montras figuro~(\ref{Figura1}.a). Ni konsideras la preterpason kun $b\neq0$ kaj la trapason kun $b=0$. Krome ni komparas niajn rezultojn al tiujn de~\cite{PaivaTeixeira2007}.

^Ci tie $x$ estas la loko de fonto je momento $t$, amba^u mezurataj per la inercia referenco sistemo $S$. La tempo $\tau$, anka^u en $S$, estas la propratempo de la restanta observanto kiam li ricevas signalon el $x$ je $t$. Kiel figuro~(\ref{Figura1}.a) evidentigas, la intertempo $\tau-t$ inter eligo kaj enigo de signalo estas $\sqrt{x^2+b^2}/c$. 

Por simpligi formulojn ni formale konsideros $c=1$ kaj $g=1$. Por aperigi la arbitrajn valorojn sufi^cas substitui 
\bea                                                       \label{cg1}
a\!\rightarrow\! ag/c^2\ , \hskip1mm A\!\rightarrow\! Ag/c^2\ , \hskip1mm b\!\rightarrow\!bg/c^2\ , \hskip1mm x\!\rightarrow\!gx/c^2\ , \hskip1mm v\!\rightarrow\!v/c\ , \hskip1mm t\rightarrow\!gt/c\ , \hskip1mm \tau\!\rightarrow\!g\tau\!/c\ , \hskip1mm \tau'\!\rightarrow\!g\tau'\!/c\ . 
\eea  
Farinte tiun simpligon kaj uzante la kondi^cojn $x(0)=-a$ kaj $v(0)=0$, la rapido $v$ kaj la loko $x$ de la fonto ^ce konstanta propra akcelo $g$ estas~\cite{PaivaTeixeira2007}
\bea                                                     \label{vtKxt}
v(t)=\frac{t}{\sqrt{1+t^2}}\ , 
\hspace{2em}
x(t)=\sqrt{1+t^2}-A\ ,
\hspace{2em}
A:=a+1\ .
\eea
^Car la fonto ne restas en $S$, tial la propratempo de eligo de signalo estas $\tau'\neq t$. Vere $\dd\tau'=\dd t/\gamma(t) = \sqrt{1-v^2(t)/c^2} \dd t$.  La integro de $\dd\tau'$ fiksante $\tau'=0$ kiam $t=0$ estas
\bea                                                     \label{ttau'}
t\!=\!\sinh\tau'\ ,
\eea
do la rapido kaj la loko de la fonto kiel funkcioj de $\tau'$ estas
\bea                                               \label{vtau'Kxtau'}
v(\tau')=\tanh\tau'\ ,
\hspace{2em}
x(\tau')=\cosh\tau'-A\ . 
\eea
Vidu ke la preterpasoj ($x=0$) okazas kun rapido $v_0:=\sqrt{A^2-1}/A$ je $\mp t_0$ (a^u $\mp \tau'_0$), kie
\bea                                                  \label{t0Ktau'0}
t_0 := \sqrt{A^2-1}\ , 
\hspace{2em}
\tau'_0 := \cosh^{-1}\!\!A\ . 
\eea 

Interesas kalkuli la rilaton inter la tempo $t$ de eligo kaj la tempo $\tau$ de enigo. El figuro~(\ref{Figura1}.a) 
\bea                                                      \label{taut}
\tau\!=t+\sqrt{x^2+b^2}
=t+\sqrt{(\sqrt{1+t^2}-A)^2+b^2}\ , 
\eea 
uzante~(\ref{vtKxt}). Figuro~(\ref{Figura1}.b) montras ke la enigo ne komencas kiam $\tau=-\infty$, sed jese kiam $\tau\!=\!-A$. La fonto eligis tiujn komencajn signalojn je $t\!=\!-\infty$ el $x\!=\!\infty$, kiam ^gia rapido estis $-c$. Observu anka^u la asimptoton $\tau\!=\!2\,t-A$ por $t\rightarrow\infty$; la fonto eligos tiujn signalojn je $t\!=\!\infty$, kaj ili enigos poste `duoble' nefinia tempo al observanto. 

\section{Dopplera efiko}                         \label{Dopplera}

\begin{figure}[t]                                               
\hfill
\epsfig{file=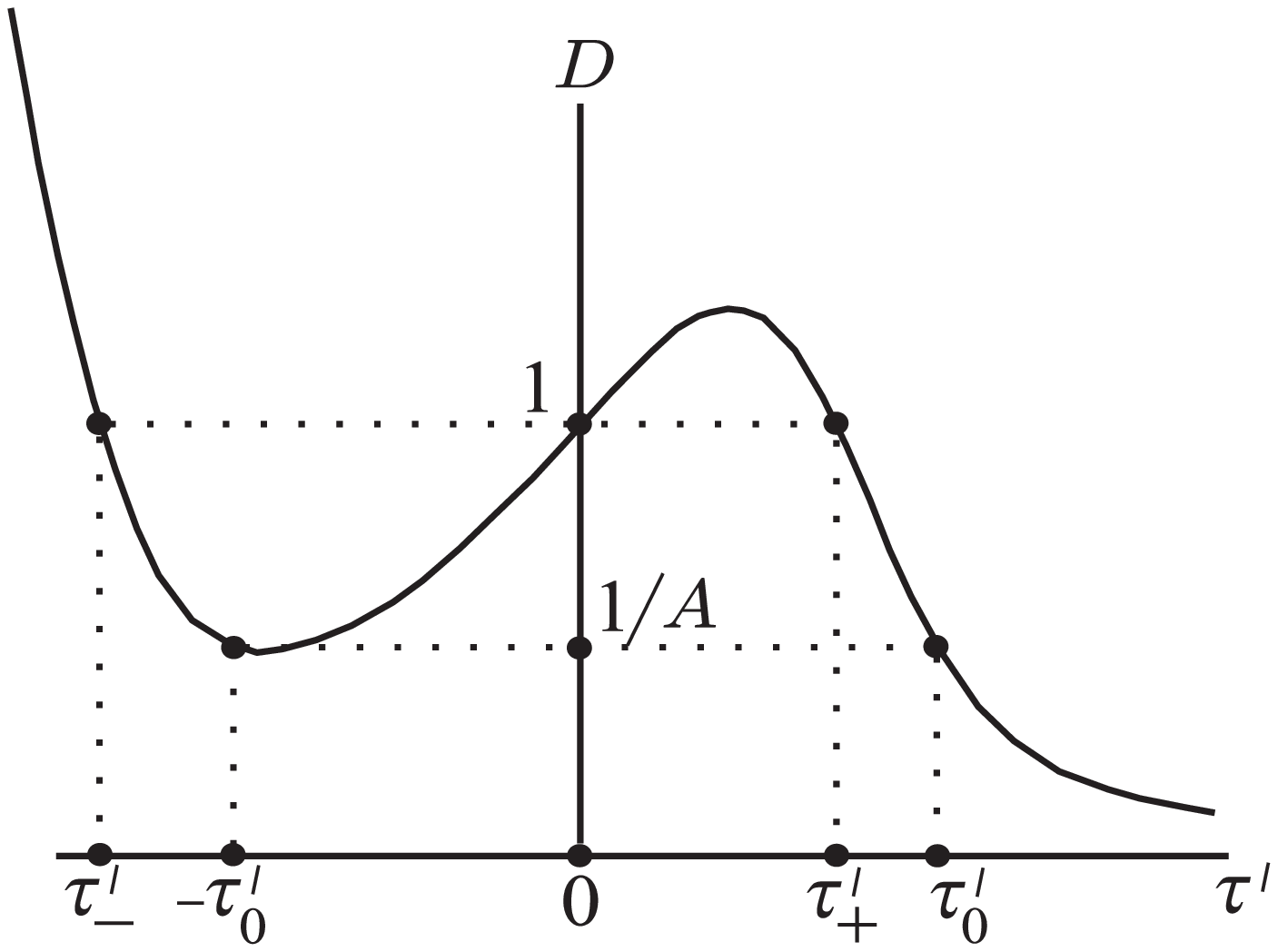,width=6cm} 
\hfill
\epsfig{file=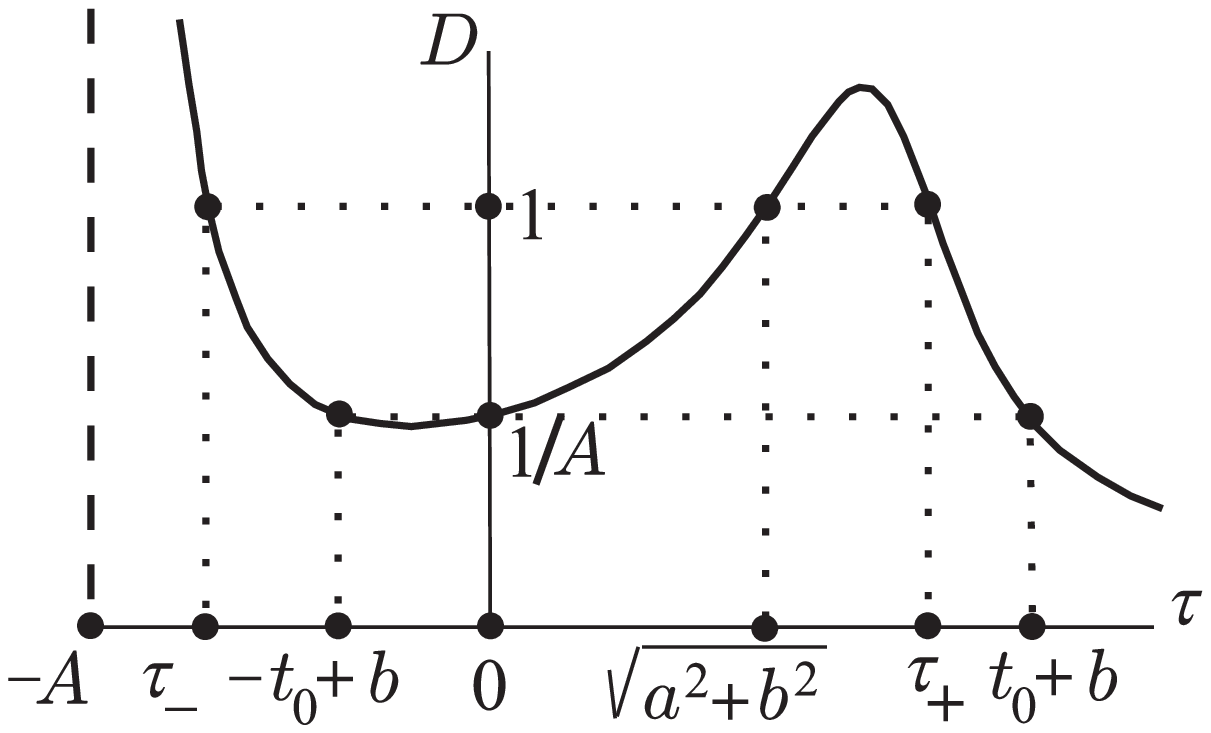,width=6cm}
\hfill \mbox{}
\caption{Doppleraj faktoroj. {\bf \ref{Figura2}.a)} $D$ kiel funkcio (\ref{Dtau'}) de la propratempo $\tau'$ de movi^ganta fonto. Estas tri momentoj en kiuj la radiado ne havos kolor-^san^gon $(D\!=\!1)$; la unua ($\tau'_-$) anta^uas al momento $-\tau'_0$ de la unua preterpaso; la dua ($\tau'\!=\!0$) okazas kiam la fonto restas ^ce $x\!=\!-a$; kaj la tria ($\tau'_+$) anta^uas al momento $\tau'_0$ de la dua preterpaso. La radiadoj ^ce la momentoj $\mp\tau'_0$ de preterpaso havas $D\!=\!1/A$, do ili estas ru^g-delokigataj. La valoroj de $\tau'_-$ kaj $\tau'_+$ estas ^ce (\ref{Dtau'1}), kaj $\tau'_0$ estas ^ce (\ref{t0Ktau'0}).
\newline
{\bf \ref{Figura2}.b)} $D$ kiel funkcio de la momento  $\tau$ de ricevo de la lumo. La unuaj signaloj estas ricevataj kiam  $\tau\!=\!-A$, kvankam ili estis eligataj detempe de $\tau\!=\!-\infty$. Ne estas kolor-^san^go je $\tau_-$, $\sqrt{a^2+b^2}$, kaj $\tau_+$. La radiado eligita en la preterpaso, $\mp t_0$ en ekv.~(\ref{t0Ktau'0}), estas observata je $\mp t_0+b$, ekv.~(\ref{taut}), kun $D=1/A$. }
\label{Figura2}
\end{figure}

\begin{figure}[t]                                               
\hfill
\epsfig{file=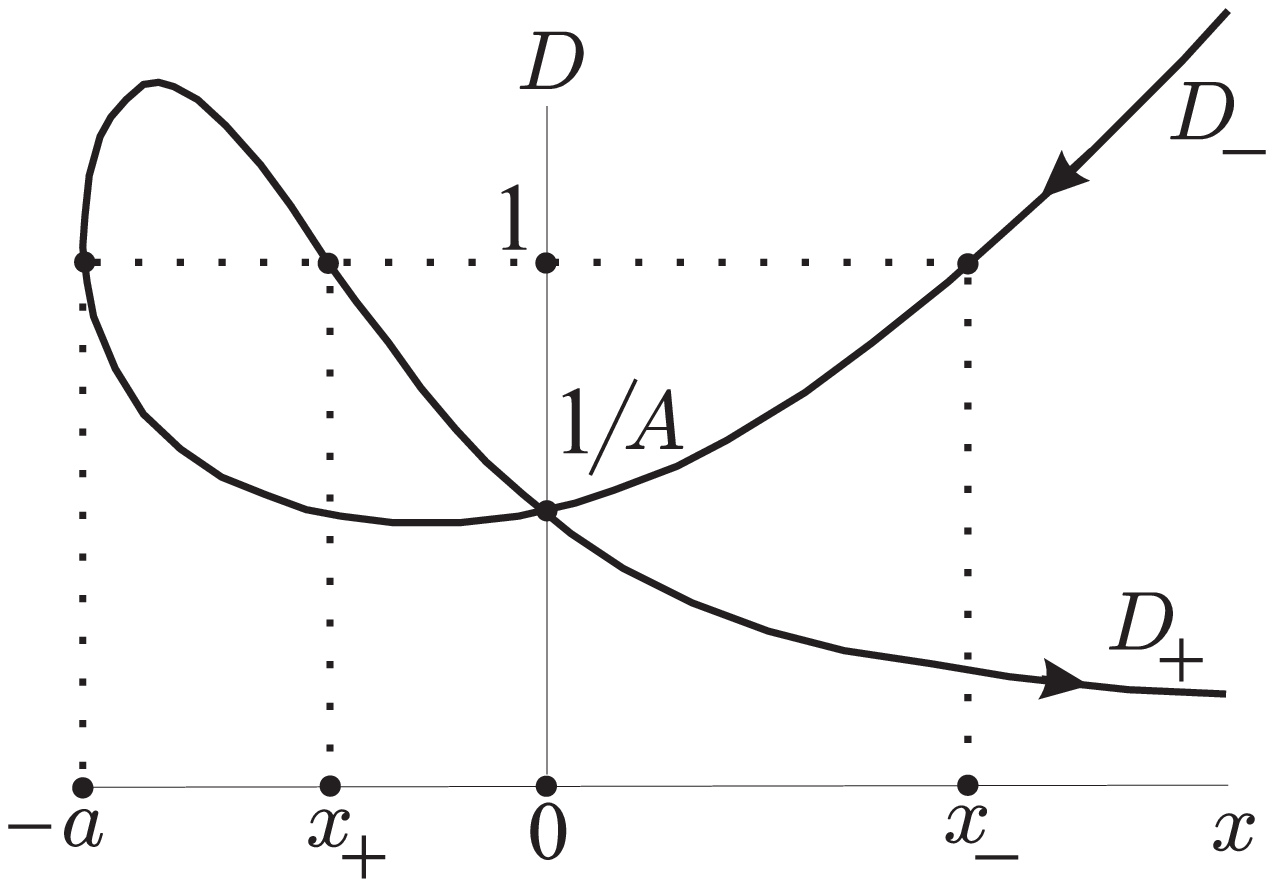,width=6cm}
\hfill
\epsfig{file=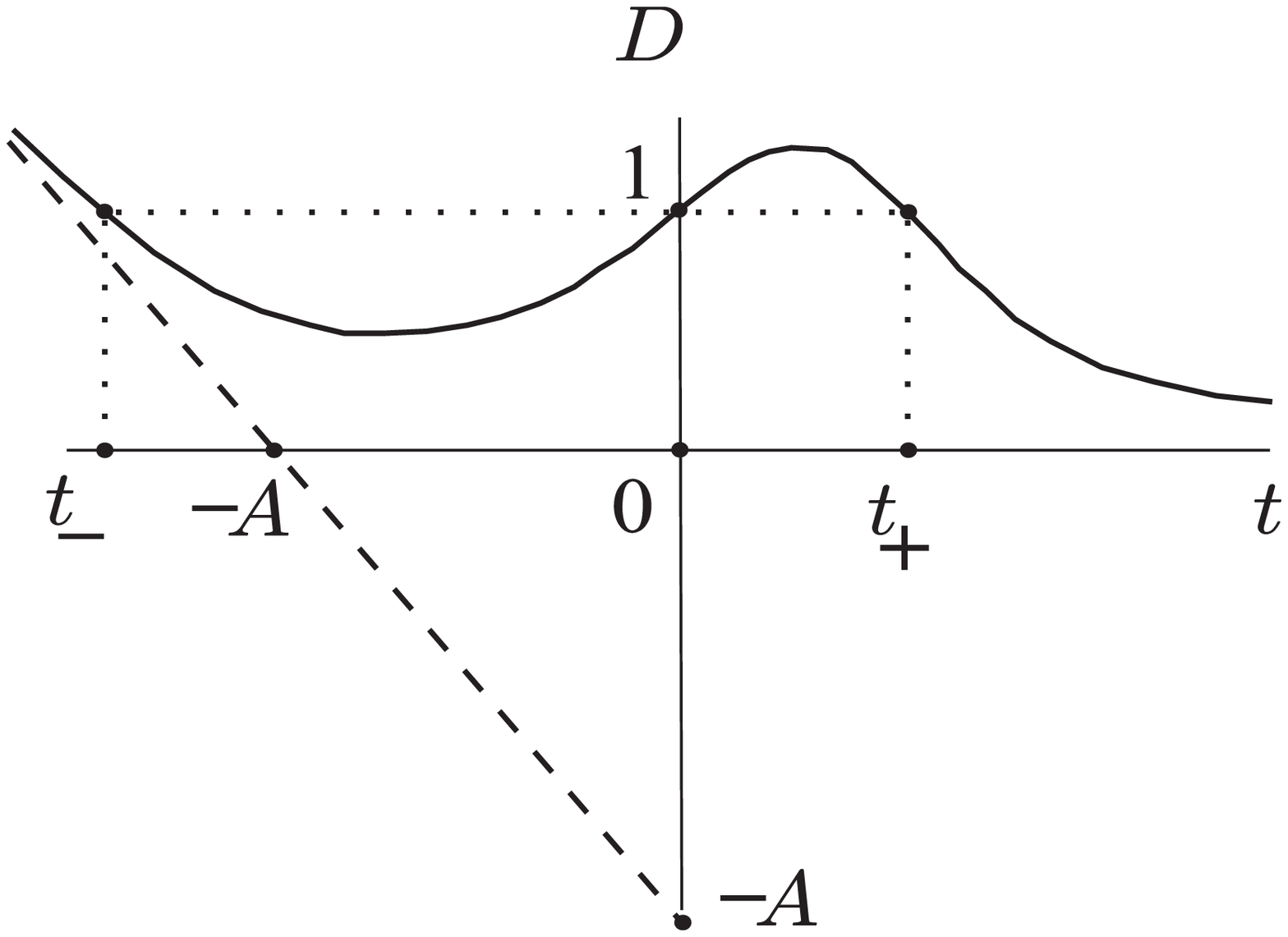,width=6cm}
\hfill \mbox{}
\caption{Doppleraj faktoroj. {\bf \ref{Figura3}.a)} $D_-$ kaj $D_+$ kiel funkcio (\ref{Dx}) de loko $x$ de lum-eligo el la movi^ganta fonto. Sagoj montras la pozitivan fluon de tempo. Radiado el $x_-$, $-a$, kaj $x_+$ ne kolor-^san^gos, $D\!=\!1$. La valoroj de $x_-$ kaj $x_+$ estas en (\ref{Dx1}). Radiadoj el preterpasoj $x\!=\!0$ havos $D\!=\!1/A$, do ili estos ru^g-delokigataj. 
\newline
{\bf \ref{Figura3}.b)} $D$ kiel funkcio (\ref{Dt}) de tempo  $t$ de referenc-sistemo $S$. Vidu ke $D\!=\!1$ kiam $t_-\!=\!\sinh\tau'_-$, $t\!=\!0$, kaj $t_+\!=\!\sinh\tau'_+$, estante $\tau'_-$ kaj $\tau'_+$ prezentataj en (\ref{Dtau'1}). Vidu anka^u la asimptoton $D\!=\!-t\!-\!A$.}
\label{Figura3}   
\end{figure}

^Car frekvenco estas la inverso de periodo, tial ni povas redifini la Doppleran faktoron kiel 
\bea                                                       \label{D}
D(\tau):=\frac{\dd\tau'}{\dd\tau} = \sqrt{1-v^2(t)}\frac{\dd t}{\dd\tau}\ .
\eea 
^Ci tie $\dd\tau'$ estas la infinitezima propra intertempo per la fonto inter eligo de du lumaj signaloj, kaj $\dd\tau$ estas la infinitezima propra intertempo per la observanto inter enigo de tiuj du lumaj signaloj. La radiko rilatas al la tempa dilato pro la movado de la fonto. Kiam $D < 1$ la Dopplera efiko estas nomata `ru^g-delokigo', kaj kiam $D>1$ ^gi estas nomata `viol-delokigo'. Ni emfazas ke estas aliaj difinoj de Dopplera faktoro~\cite{Roth}.

Sekve ni kalkulas la Doppleran faktoron kiel funkcio de $\tau'$, $\tau$, $x$ kaj de $t$. Ni komencas kun $D(\tau')$. Uzante la difinon~(\ref{D}) en la formo $D=(\dd\tau/\dd\tau')^{-1}$, derivante (\ref{taut}) kaj uzante $t(\tau')$ kaj $x(\tau')$ el~(\ref{ttau'}) kaj~(\ref{vtau'Kxtau'}), la faktoro esti^gas
\bea                                                    \label{Dtau'}
D(\tau')=\left(\cosh\tau'+\frac{(\cosh\tau'-A)\sinh\tau'}{\sqrt{(\cosh\tau'-A)^2+b^2}}\right)^{-1}\ .
\eea 
Figuro~(\ref{Figura2}.a) montras ke ne estas Dopplera efiko ($D=1$) trifoje:
\bea                                                   \label{Dtau'1}
\tau'_-:=-2\sinh^{-1}\frac{1}{4}(\sqrt{b^2+8a}+b)\ , \hskip3mm \tau'=0\ , \hskip3mm \tau'_+:=2\sinh^{-1}\frac{1}{4}(\sqrt{b^2+8a}-b)\ .
\eea 

Nun ni kalkulas $D$ kiel funkcio de la propratempo $\tau$ de enigo de lumo al observanto. Uzante~(\ref{ttau'}) kaj~(\ref{taut}) ni havigas
\bea                                                     \label{sinh}
\sinh\tau'=\frac{1}{2(\tau^2-A^2)}\left(\tau(\tau^2-C^2)+A\sqrt{(\tau^2-C^2)^2+4(\tau^2-A^2)}\,\right)\ , \hskip3mm C^2:=A^2+b^2+1\ ;
\eea 
substituante tiun kaj $\cosh\tau'\!=\!\sqrt{1+\sinh^2\tau'}$ en ekvacio~(\ref{Dtau'}) ni havigas la deziratan Doppleran faktoron $D(\tau)$. Figuro~(\ref{Figura2}.b) montras tiun funkcion.

Nun ni kalkulu la Doppleran faktoron kiel funkcio de la loko $x$ de signal-eligo. Uzante~(\ref{vtKxt}), (\ref{ttau'}) kaj~(\ref{vtau'Kxtau'}) en~(\ref{Dtau'}), vidu ke 
\bea                                                       \label{Dx}
D_\epsilon(x)=\left(x+A+\frac{\epsilon x\sqrt{(x+A)^2-1}}{\sqrt{x^2+b^2}}\right)^{-1}\ , \hskip3mm \epsilon:=|t|/t. 
\eea
Figuro~(\ref{Figura3}.a) montras la du funkciojn $D_-(x)$ kaj $D_+(x)$. Observu la tri lokojn el kiuj la lumo eligita ne kolor-^san^gos:  
\bea                                                      \label{Dx1}
x_-:=\frac{b}{4}(\sqrt{b^2+8a}+b)\ , 
\hspace{1ex} 
x=-a \ ,
\hspace{1ex} 
x_+:=-\frac{b}{4}(\sqrt{b^2+8a}-b)\ .
\eea

Fine ni montras $D$ kiel funkcio de la inercia tempo $t$ de eligo:  
\bea                                                       \label{Dt}
D(t)=\left(\sqrt{1+t^2}+\frac{tx}{\sqrt{x^2+b^2}}\right)^{-1}\ , \hskip3mm x(t)=\sqrt{1+t^2}-A\ .
\eea  
Figuro~(\ref{Figura3}.b) montras tiun funkcion. Observu la asimptoton $D\!=-t\!-\!A$ kiam $t\rightarrow-\infty$. 
 
\section{Trapaso}                                 \label{trapaso}

\begin{figure}[t]                                             
\hfill
\epsfig{file=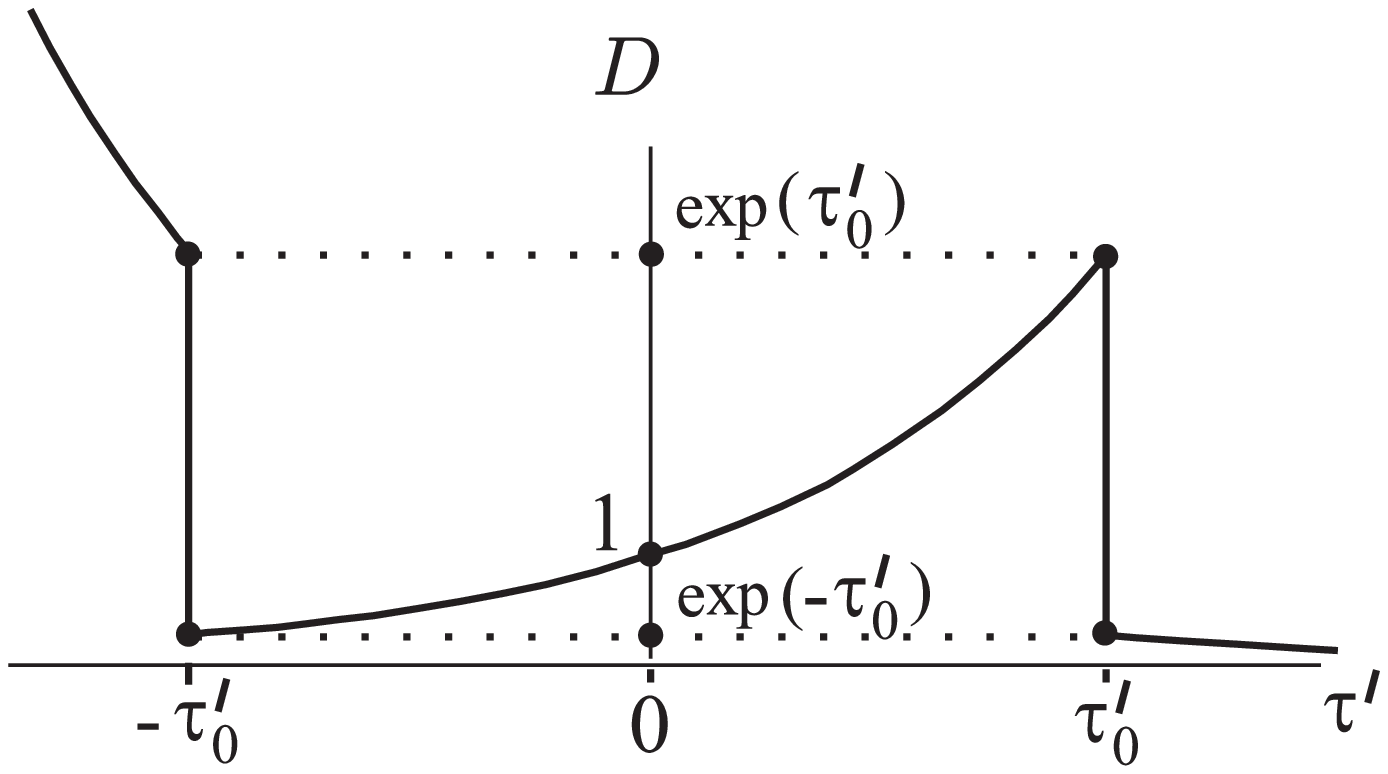,width=8cm}
\hfill
\epsfig{file=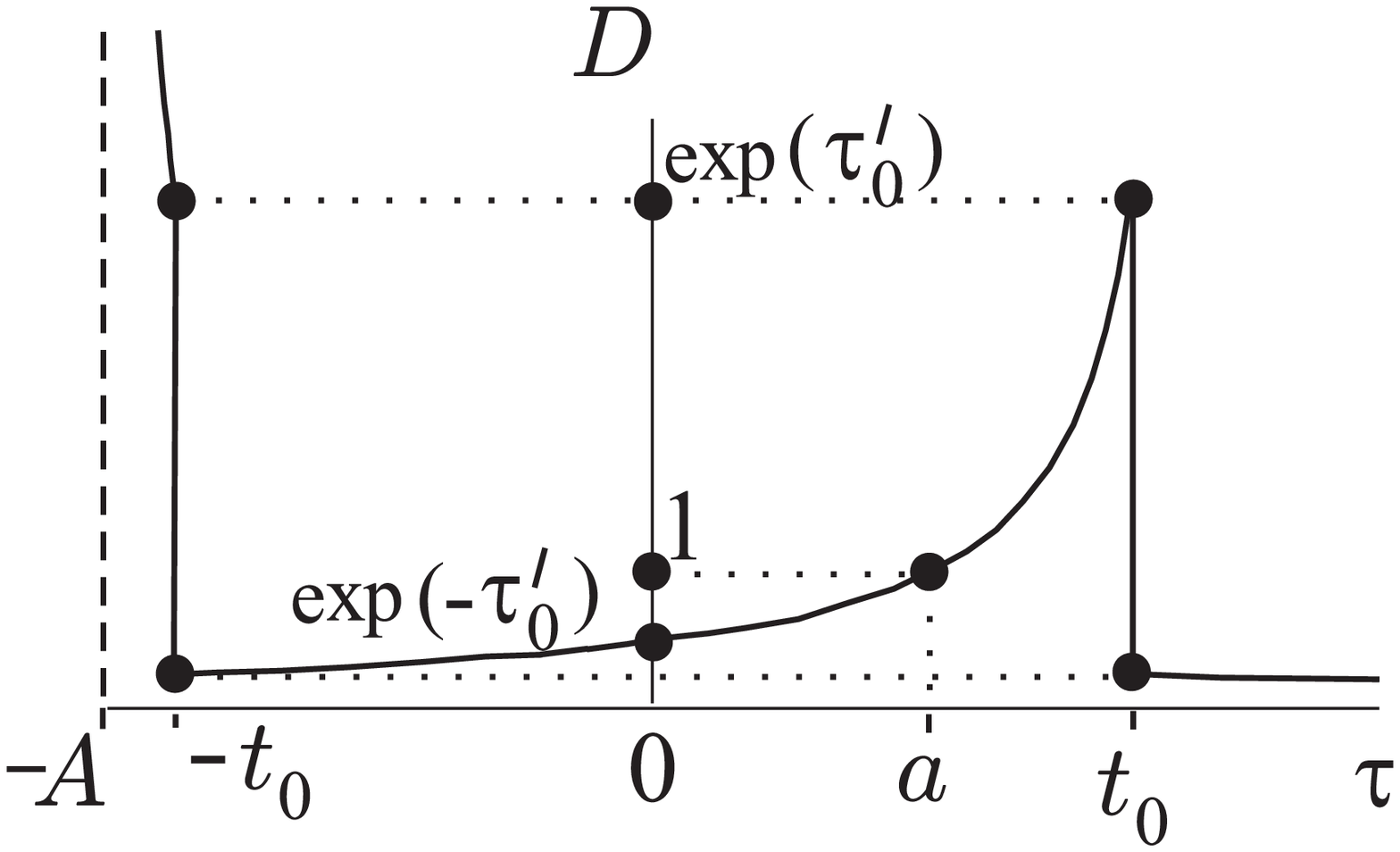,width=6cm}
\hfill \mbox{}
\caption{Doppleraj faktoroj kun $b=0$. {\bf \ref{Figura4}.a)} $D$ kiel eksponencialaj funkcioj (\ref{Dtau'b0}) de propratempo $\tau'$ de movi^ganta fonto. Estas nekontinueco de $D$ en momentoj $\mp\tau_{0}^{'}$\ , prezentataj en (\ref{t0Ktau'0}).
\newline
{\bf \ref{Figura4}.b)} $D$ kiel funkcio (\ref{Dtaub0}) de propratempo $\tau$ de observanto. Estas nekontinueco de $D$ en momentoj $\mp t_0$ de trapaso. Kiam  $\tau\!=\!a$ ne estas kolor-^san^go. Vidu la asimptoton $\tau\!=\!-A$; do la observanto ricevas la unuajn signalojn en ^ci tiu momento, nefinie viol-delokigataj. La valoro de $t_0$ kaj de $\tau'_0$  estas en (\ref{t0Ktau'0}).}
\label{Figura4} 
\end{figure}  

\begin{figure}[t]                                            
\hfill
\epsfig{file=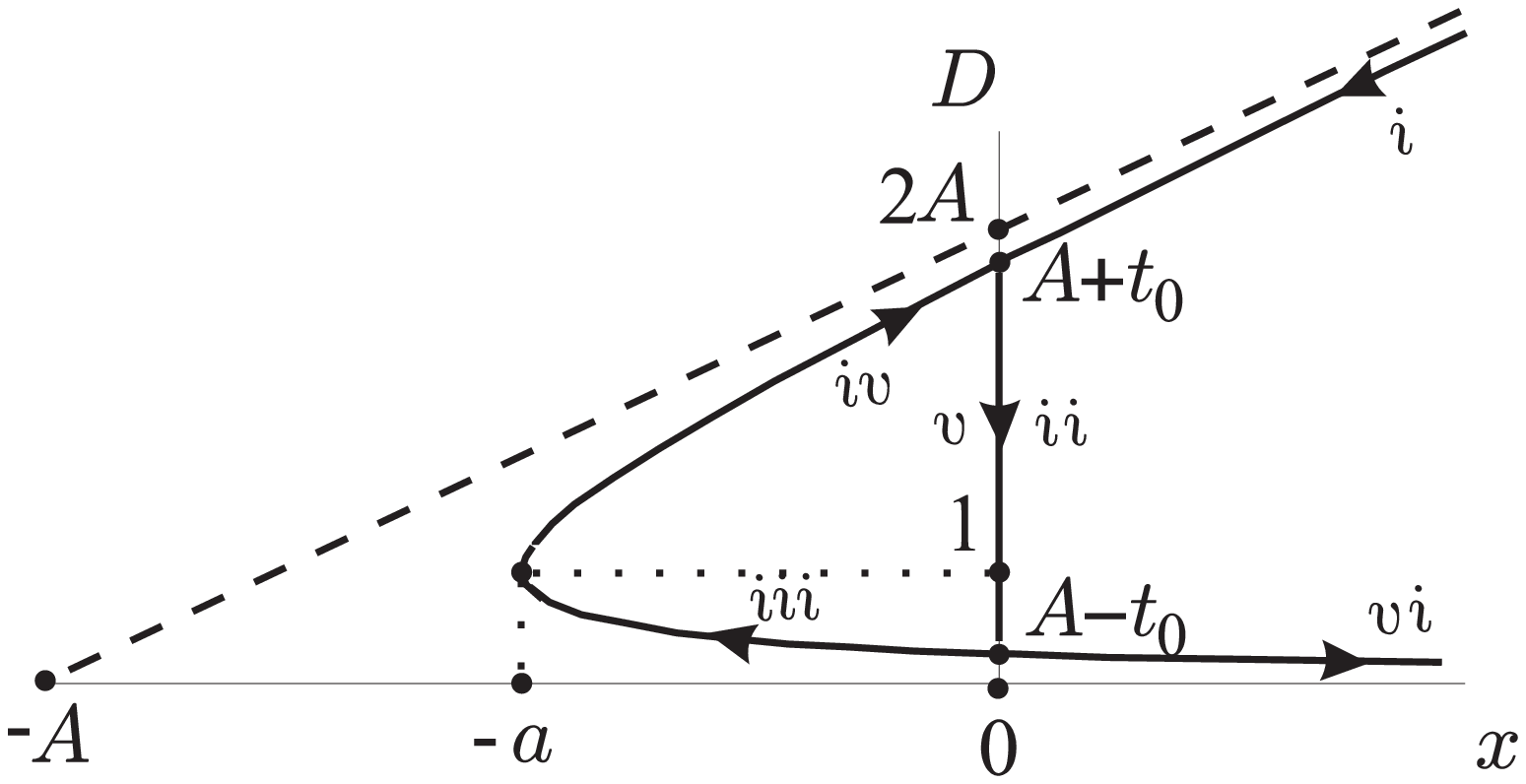,width=8cm}
\hfill
\epsfig{file=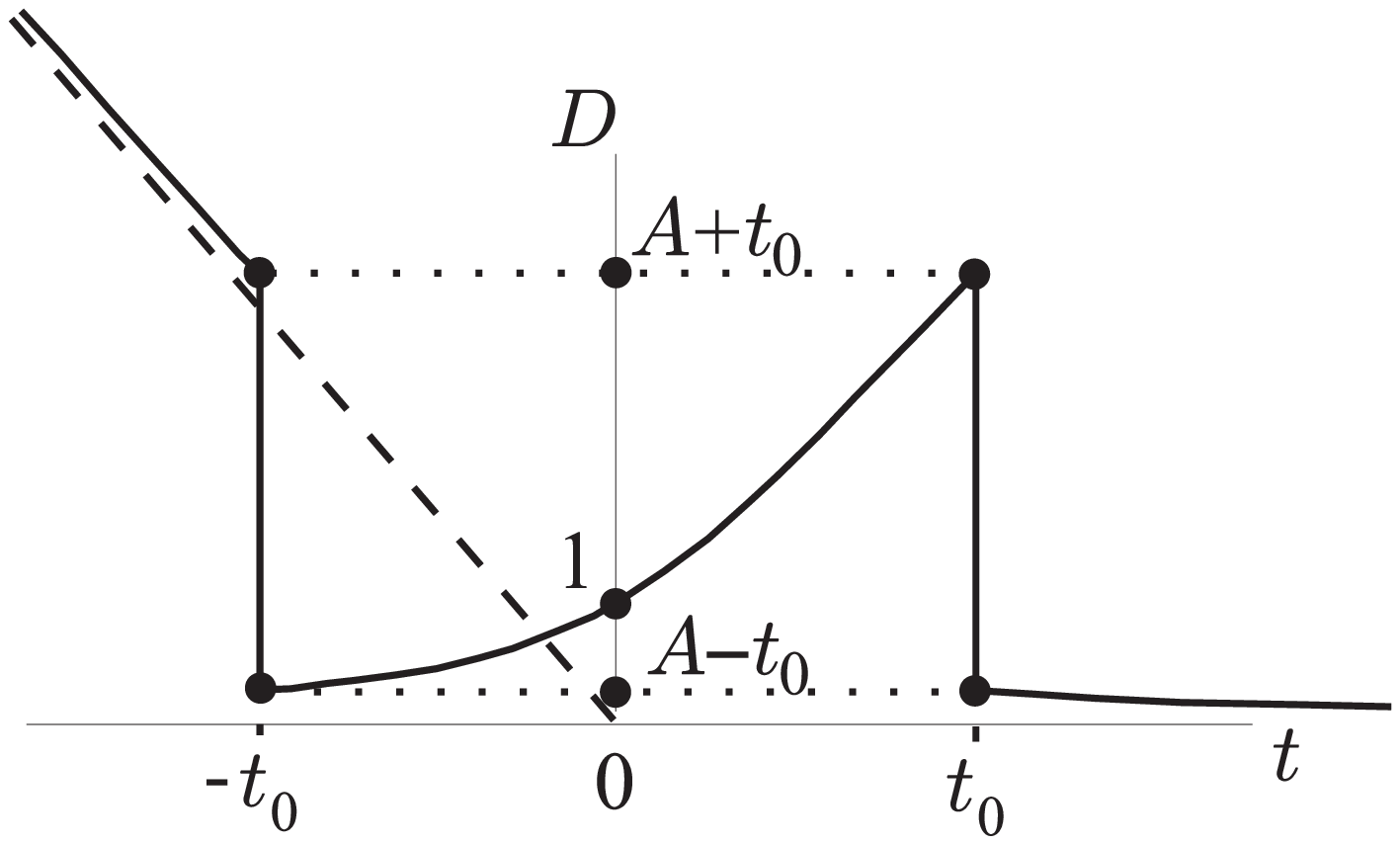,width=8cm}
\hfill \mbox{}
\caption{Doppleraj faktoroj kun $b=0$. {\bf \ref{Figura5}.a)} $D$ kiel funkcio (\ref{Dxb0}) de la loko $x$ de lum-eligo. Sagoj montras la pozitivan fluon de la tempo. Vidu la asimptoton  $D\!=\!2(x+A)$. Estas nekontinueco de $D$ en $x\!=\!0$. La valoro de $\tau_{0}'$ estas en (\ref{t0Ktau'0}).
\newline
{\bf \ref{Figura5}.b)} $D$ kiel funkcio (\ref{Dtb0}) de $t$. Estos nekontinueco de $D$ en lum-eligoj kiam $\mp t_0$, (\ref{t0Ktau'0}). Vidu asimptoton $D\!=\!-2t$. La valoro de $\tau_{0}'$ estas en (\ref{t0Ktau'0}).} 
\label{Figura5} 
\end{figure}  

Nun ni studas la interesan okazon ^ce $b\!=\!0$. Anstata^u preterpasi, la fonto pasas \emph{tra} la observanto. Tio generas nekontinuecon $\Delta D$ je la Dopplera faktoro. Por havigi $D$ kiel funkcio de $\tau'$, $\tau$, $x$ kaj $t$, ni faras $b=0$ en ekvacioj~(\ref{Dtau'}), (\ref{sinh}), (\ref{Dx}), (\ref{Dt}), respektive. Pri $D(\tau')$ ni havigas
\bea                                                  \label{Dtau'b0}
D(\tau')={\rm exp}(-\epsilon_x\tau')\ ,
\eea 
kie $\epsilon_x$ estas la signumo de $x$. Estas $\epsilon_x\!=\!+1$ kiam $\tau'\!<\!-\tau'_0$ kaj $\tau'\!>\!\tau'_0$, kaj estas  $\epsilon_x\!=\!-1$ kiam $-\tau'_0\!<\tau'\!<\!\tau'_0$. Vidu figuron~(\ref{Figura4}.a). 

Pri $D(\tau)$, figuro~(\ref{Figura4}.b) montras la funkcion
\bea                                                   \label{Dtaub0}
D(\tau) = \left\{
\begin{array}{lll} 
(A+\tau)^{-1}\ , & -A<\tau<-t_0\ , & \tau>t_0\ , 
\\ \mbox{} \\ 
(A-\tau)^{-1}\ ,  & -t_0<\tau<t_0\ . 
\end{array}
\right.
\eea

Pri $D(x)$ ni havigas
\bea                                                     \label{Dxb0}
D_\epsilon(x)=x+A-\epsilon\epsilon_x\sqrt{(x+A)^2-1}\ . 
\eea 
Vidu figuron~(\ref{Figura5}.a), observante la tempan sinsekvon $i$, $ii$, $iii$, $iv$, $v$, $vi$. 

Fine, pri $D(t)$,
\bea                                                     \label{Dtb0}
D_x(t)=\sqrt{1+t^2}-t\epsilon_x\ , 
\eea 
kie $\epsilon_x$ estas la signumo de $x$. Estas $\epsilon_x\!=\!+1$ kiam $t\!<\!-t_0$ kaj $t\!>\!t_0$, kaj estas $\epsilon_x\!=\!-1$ kiam $-t_0\!<\!t\!<\!t_0$. Vidu figuron~(\ref{Figura5}.b).  

La nekontinueco $\Delta D$ de Dopplera faktoro $D$ estas facile kalkulata ekde la anta^uaj rezultoj:
\bea                                                  \label{descont}
\Delta D=\sqrt{\frac{1+v_0}{1-v_0}}-\sqrt{\frac{1-v_0}{1+v_0}}=\frac{2v_0}{\sqrt{1-v_0^2}}=2\sqrt{A^2-1}=2\,\sinh\tau'_0=2\,t_0\ .
\eea

\section{Komentoj}                                               

Ni konsideris restantan observanton ^ce inercia sistemo $S$ kaj movi^gantan fonton, kiel figuro~(\ref{Figura1}.a). Ni montris, vidu figuron~(\ref{Figura3}.a), ke lumo eligita el $x\!=\!0$ (minimuma distanco al observanto, ^ce $S$) estas enigota ru^g-delokigate. Tamen, en la anta^ua artikolo~\cite{PaivaTeixeira2007} (vidu ^gian figuron~1), kie la fonto restas ^ce inercia sistemo $\Sigma$ kaj la observanto movi^gas,  lumo enigita kiam la observanto estas en $x=0$ estas vidata kun viol-delokigo. Tio malsamo estas natura en special-relativeco, kaj ni plieksplikos tion en estonta artikolo de revizio; vidu anka^u~\cite{PaivaTeixeira2006}. 

^Ci tie ni vidis en figuroj (\ref{Figura2}.b) kaj (\ref{Figura4}.b), ke la observanto ricevas signalojn el la fonto nur poste sia finia  propratempo $\tau\!=\!-A$. ^Ci tiu kontrastas kun ekvacioj (19) kaj (17) de \cite{PaivaTeixeira2007}, kiuj indikas ke la observanto ricevas signalojn je ^ciuj momentoj $-\infty\!<\!\tau\!<\!\infty$. La kialo de tiu malsameco estas, ke ^ce \cite{PaivaTeixeira2007} la tempo $\tau$ estas propratempo de movi^ganta observanto, kvankam en ^ci tiu artikolo la tempo $\tau$ estas propratempo de restanta observanto. La fluoj de la du tempoj estas tre malsamaj. 

En estonta artikolo ni prezentos sistemojn kies fonto kaj observanto amba^u estas akcelataj. Kelkaj el ^giaj rezultoj estas mirindaj.

\end{document}